\documentclass[final,5p,times,twocolumn]{elsarticle}

\usepackage{amssymb}
\usepackage{amsmath}
\usepackage{upgreek}

\usepackage{booktabs}

\usepackage{tikz}
\usetikzlibrary{arrows.meta,calc}

\usepackage{hyperref}

\journal{Nuclear Instruments and Methods A}

\begin{document}

\begin{frontmatter}



\title{The scintillating fibre beam monitor for the CHNet-MAXI experiment at ISIS: characterisation of the light collection and detection system}



\author[unipv,infnpv]{L. Malliolio$^{\dag,}$}
\author[unipv,infnpv]{R. Rossini$^{\dag,}$\corref{cor2}}
\author[nuclear]{A. Abba}
\author[infnmb]{R. Bertoni}
\author[isis]{S. Biswas}
\author[infnmb]{M. Cataldo}
\author[infnmb]{G. Ceruti}
\author[infnpv]{C. De Vecchi}
\author[unimb,infnmb]{D. Di Martino}
\author[isis]{A.D. Hillier}
\author[isis]{J.S. Lord}
\author[unimb,infnmb,isis]{G. Marcucci}
\author[unipv,infnpv]{A. Menegolli}
\author[cnao,infnmi]{A. Mereghetti}
\author[uniss,infnca]{P. Monza}
\author[dista,infnpv]{M. Musa}
\author[unipi,infnpisa]{P. Oliva}
\author[dista]{F. Palenzona}
\author[infnmb]{M. Perego}
\author[isis]{D.E. Pooley}
\author[infnpv]{M.C. Prata}
\author[cnao,infnpv]{M. Pullia}
\author[isis]{L. Quintieri}
\author[dista,infnpv]{M.P. Riccardi}
\author[uniss,infnca]{A. Roggio}
\author[infnpv]{M. Rossella}
\author[cnao,infnmi]{S. Savazzi}
\author[uniss,infnca]{V. Sipala}
\author[infnmb]{E.S. Vallazza}
\author[cnao]{C. Viviani}
\author[infnmb]{M. Clemenza}

\cortext[cor2]{Corresponding author: riccardo.rossini@infn.it. (R. Rossini) \\
\indent \dag \ These authors contributed equally to this work.}

\affiliation[unipv]{
    organization={Department of Physics "A. Volta", University of Pavia},
    city={Pavia},
    country={Italy}
}

\affiliation[infnpv]{
    organization={INFN sezione di Pavia},
    city={Pavia},
    country={Italy}
}

\affiliation[nuclear]{
    organization={Nuclear Instruments S.r.l.},
    addressline={Lambrugo (CO)},
    country={Italy}
}

\affiliation[infnmb]{
    organization={INFN sezione di Milano-Bicocca},
    city={Milan},
    country={Italy}
}

\affiliation[isis]{
    organization={ISIS Neutron and Muon Source, Science and Technology Facilities Council},
    city={Didcot},
    country={United Kingdom}
}

\affiliation[unimb]{
    organization={Department of Physics "G. Occhialini", University of Milano-Bicocca},
    city={Milan},
    country={Italy}
}

\affiliation[cnao]{
    organization= {Centro Nazionale di Adroterapia Oncologica (CNAO)},
    city = {Pavia},
    country={Italy}
}

\affiliation[infnmi]{
    organization={INFN sezione di Milano},
    city={Milan},
    country={Italy}
}

\affiliation[uniss]{
    organization={Department of Chemical, Physical, Mathematical and Natural Sciences, University of Sassari},
    city={Sassari},
    country={Italy}
}

\affiliation[infnca]{
    organization={INFN sezione di Cagliari},
    city={Cagliari},
    country={Italy}
}

\affiliation[dista]{
    organization={Department of Earth and Environmental Sciences, University of Pavia},
    city={Pavia},
    country={Italy}
}

\affiliation[unipi]{
    organization={Department of Physics "E. Fermi", University of Pisa},
    city={Pisa},
    country={Italy}
}

\affiliation[infnpisa]{
    organization={INFN sezione di Pisa},
    city={Pisa},
    country={Italy}
}

\begin{abstract}
The CHNet-MAXI experiment at ISIS aims at a non-destructive isotopic characterisation of lead samples, based on muonic atom formation. A custom scintillation-based detector is used for the reconstruction of the time and space distributions of the low-energy muon beam employed in the experiment. The beam monitor consists of two orthogonal planes of 10 cm-long, 3 mm × 3 mm square scintillating fibres, totalling 64 fibres, read out by four arrays of 16 Hamamatsu S14160-3015PS SiPMs each. The optical performance of the detector has been characterised, focusing on the SiPM behaviour, and also checking and correcting the uniformity of the light response across channels with a $^{60}$Co source. A first detector test with protons at the CNAO-XPR has been carried out. The results of the Geant4 simulation of the detector, fundamental for the detector operation and data analysis, are also presented. All these activities are preliminary to the final detector operation in the CHNet-MAXI experiment at the RIKEN-RAL Port4 decay muon beamline at ISIS. 


\end{abstract}




\end{frontmatter}



\section{Introduction}
    The CHNet-MAXI (Cultural Heritage Network - Muonic Atom X-ray spectroscopy for Isotopic analysis) experiment aims to develop a non-invasive technique (Muonic Atom X-ray Emission Spectrometry, $\upmu$-XES) to determine the Pb isotopic ratios in ancient artefacts \cite{ninomiya}. The stable isotopes of lead are $^{204}$Pb, $^{206}$Pb, $^{207}$Pb, and $^{208}$Pb. Since $^{204}$Pb is primordial, and $^{206}$Pb, $^{207}$Pb, $^{208}$Pb are at the end of radioactive chains, their ratio provides important information about the provenance of lead-containing artefacts.
    This experiment is funded by the INFN CSN5 as part of the INFN Cultural Heritage Network (CHNet)\cite{Mazzinghi2024X-ray}.
    
    CHNet-MAXI is centred on detecting the X and $\gamma$ rays emitted after muonic atom formation, obtained by irradiating the samples with a beam of $\upmu^-$, similarly to what is done for muonic atom elemental analysis\cite{tandem}. The activity is being carried out at the RIKEN-RAL Port4 muon facility at the ISIS Neutron and Muon source, part of the Rutherford Appleton Laboratory (RAL), Didcot  (UK)\cite{RIKEN}. The X/$\gamma$ rays are detected through a set of High Purity Germanium (HPGe) detectors and innovative scintillation detectors\cite{monza:2026}.

    A muon beam monitor has been developed\cite{hodomaxi}, based on a mesh of scintillating fibres read-out by Silicon PhotoMultipliers (SiPMs), to allow sample centring, live beam diagnostics, and data normalisation. This design follows the experience gained with the FAMU\cite{FAMU1, FAMU2} experiment beam monitor development\cite{Hodo1, Hodo2}. In particular, the detector simulation, characterisation, and operation follow the procedure previously optimised\cite{rossini2024hodo}. The beam monitor is essential both during the experiment, to centre and focus the beam on the sample, and for data analysis, to normalise the spectra over the number of muons collected during each experimental run. Data normalisation over the number of muons is a crucial step towards a quantitative and accurate method. 

    \begin{figure*}
        \centering
        \includegraphics[width=0.375\linewidth]{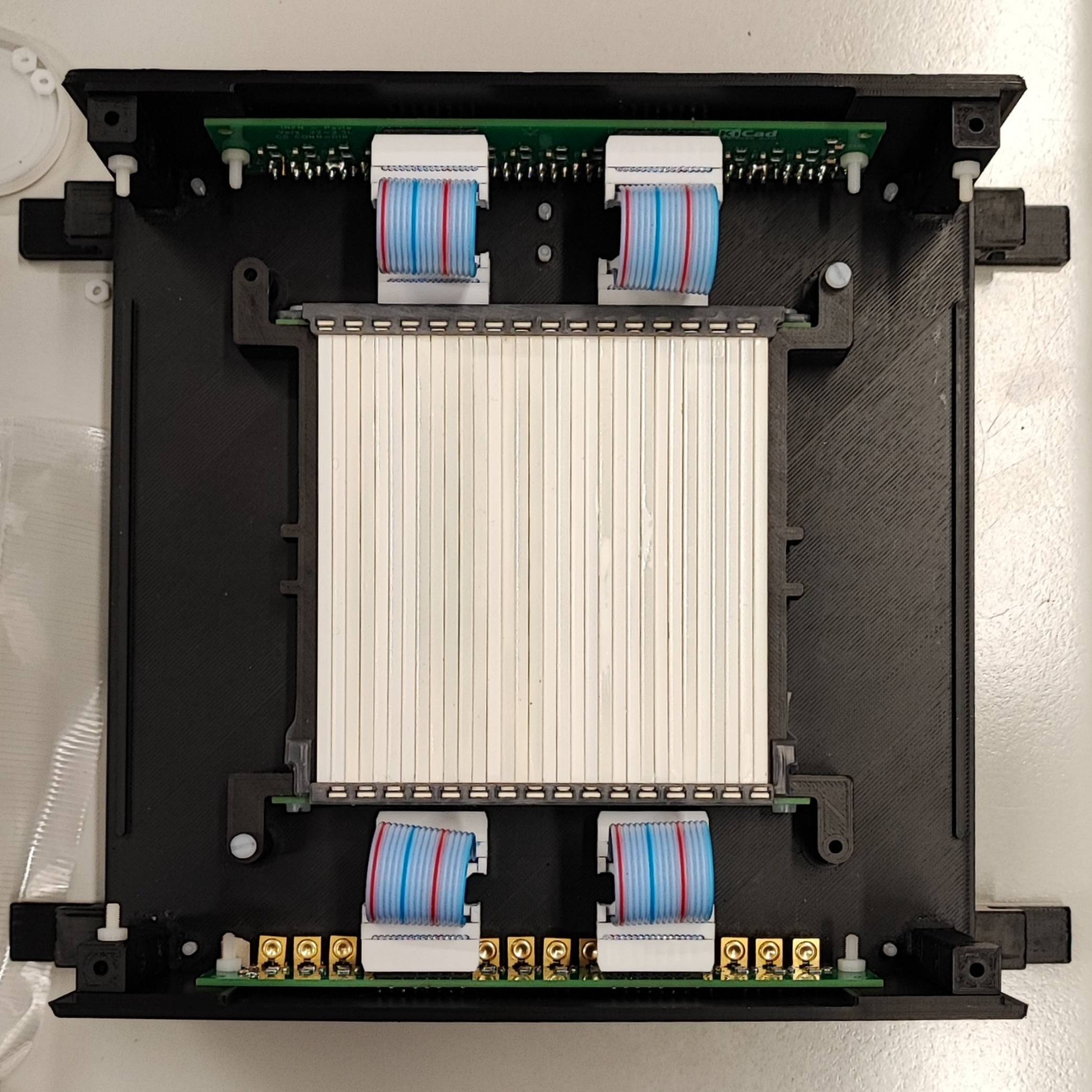}
        \includegraphics[width=0.5\linewidth]{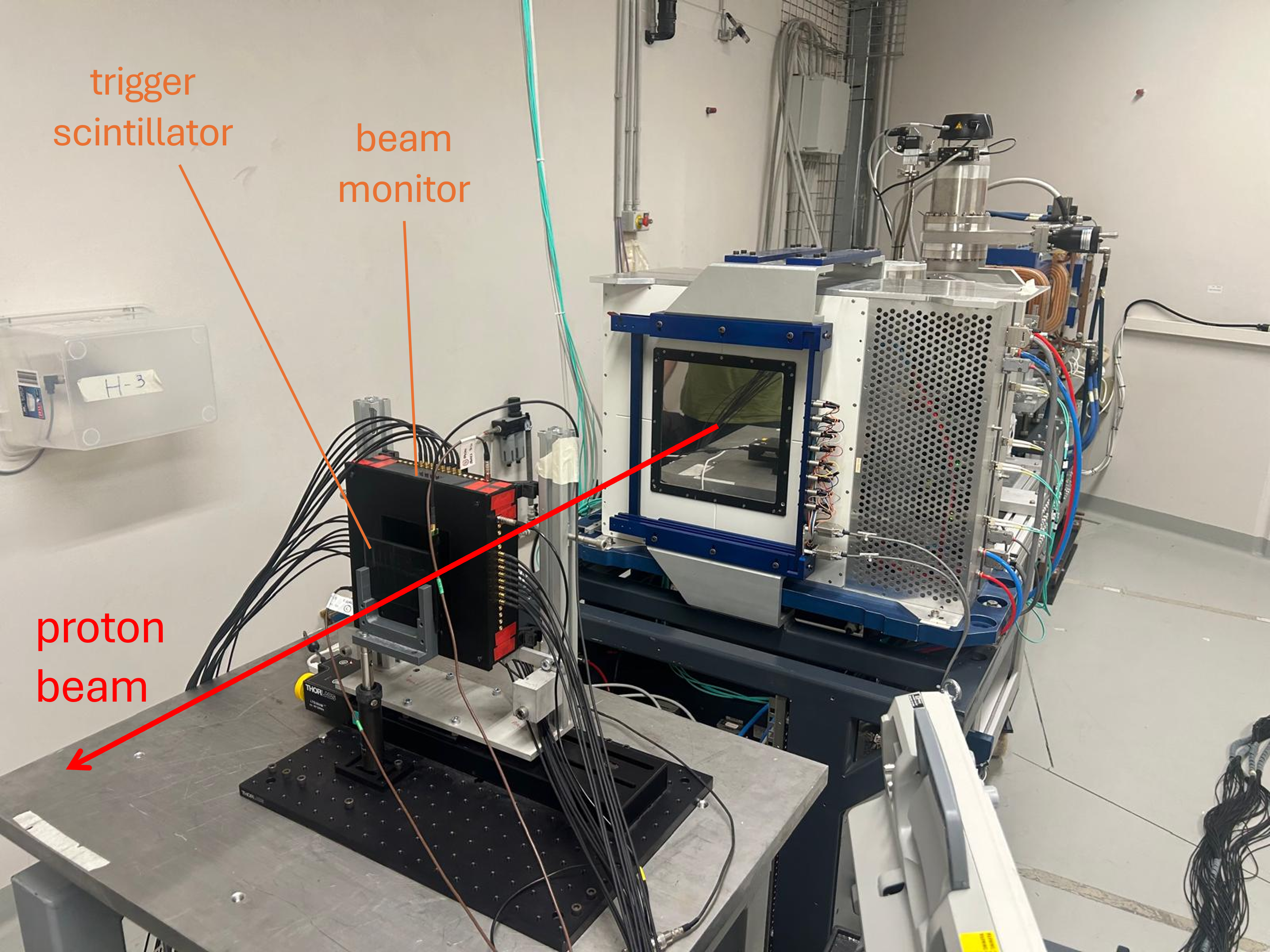}
        \caption{CHNet-MAXI beam monitor: view of one of the two scintillating fibre planes (left), and view of the detector mounted on the XPR beamline with a downstream scintillator for DAQ triggering (right).}
        \label{fig:hodo}
    \end{figure*}

    This paper illustrates the first tests with the detector at the CNAO-XPR beamline (Pavia, Italy) with a beam of 70-150 MeV protons, the channel equalisation procedure, and the detector simulation. These activities are all preliminary to the exposure to the RIKEN-RAL muon beam for absolute calibration and final detector deployment in the CHNet-MAXI experiment. 

\section{The CHNet-MAXI beam monitor}\label{sec:hodo}
    The CHNet-MAXI experiment has been equipped with a muon beam monitor\cite{hodomaxi} composed of two planes of 32 square 3 mm juxtaposed fibres, one on the X and one on the Y plane, both perpendicular to the beam direction. The fibres have been cut at a length of 10.43 cm, polished on the sides, and sprayed with a layer of $\sim 15 \ \upmu$m of diffusive TiO$_2$ varnish, as an optical coating. The fibres are then optically coupled to Hamamatsu S14160-3015PS SiPMs at either sides, mounted on boards of 16 SiPMs each, designed at INFN Pavia. All 64 SiPMs have been tested, and an average breakdown voltage of $V_{bd} = 39.21(3)$ V has been measured. The boards allow for a single bias supply for all 16 SiPMs, and to fan out the signals from each SiPM through an MCX connector. 

    The device is then enclosed into a custom 3D-printed PLA case, with 0.5 mm thick beam entrance and exit windows. One of the two fibre planes composing the detector, connected to the printed circuit boards, is shown on the left panel in Figure \ref{fig:hodo}.

    \section{Detector simulation}\label{sec:sim}

    \begin{figure}[h]
        \centering
        \includegraphics[width=0.6\linewidth]{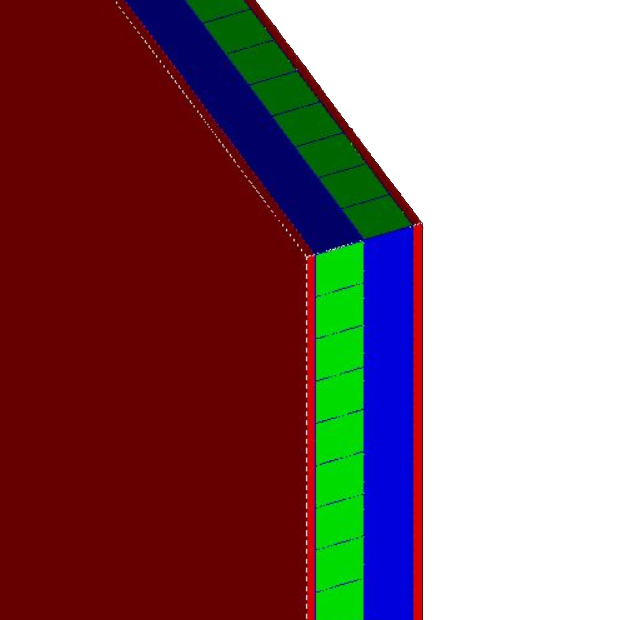}
        \caption{Close-up detail of the geometry simulated in Geant4 showing the entrance/exit windows (red), the scintillating fibres (green) and the fibre coatings (blue).}
        \label{fig:g4}
    \end{figure}

    \begin{figure*}[p]
        \centering
        \includegraphics[width=0.49\linewidth]{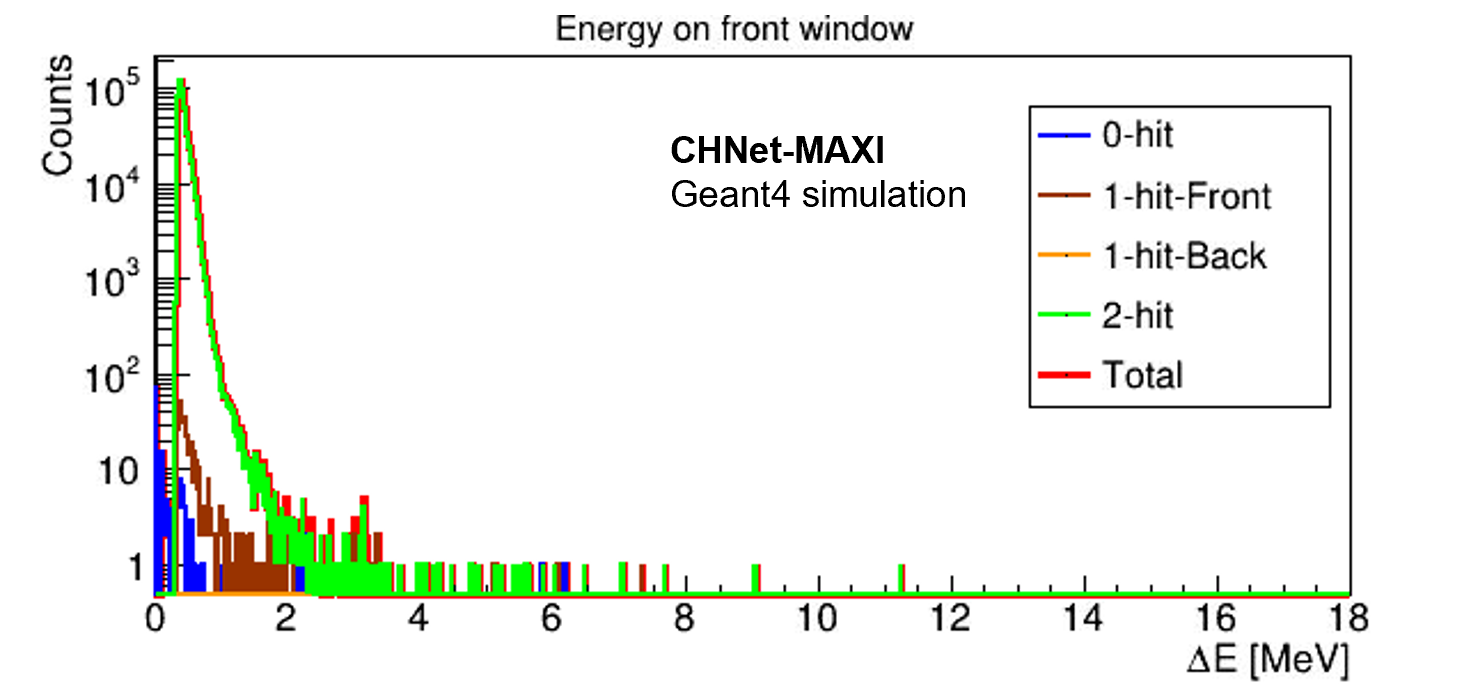}
        \includegraphics[width=0.49\linewidth]{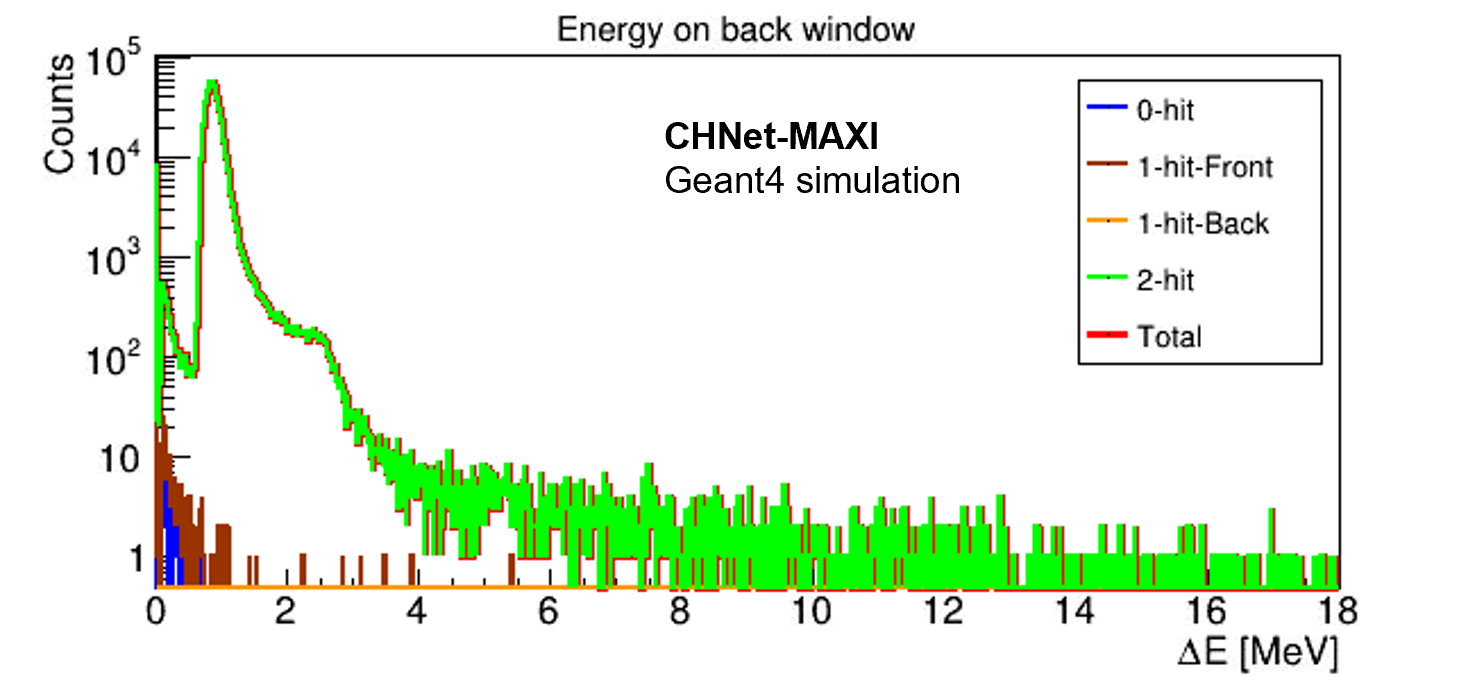}
        \caption{Simulated energy deposit on front (left) and back (right) window for 49.5 MeV/c single $\upmu^-$ interacting with 0, 1 or 2 planes of fibres ($10^6$ simulated single muons).}
        \label{fig:sim:Ew}
    \end{figure*}
    \begin{figure*}[p]
        \centering
        \includegraphics[width=0.49\linewidth]{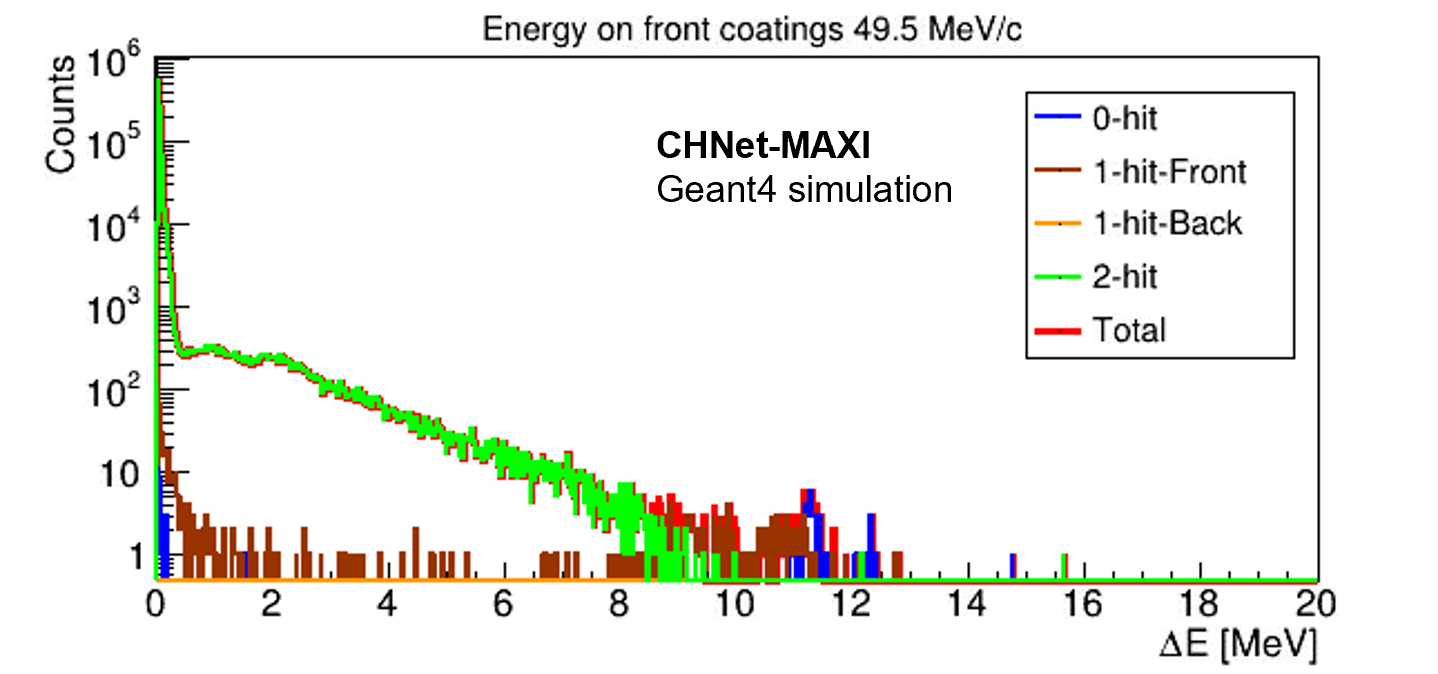}
        \includegraphics[width=0.49\linewidth]{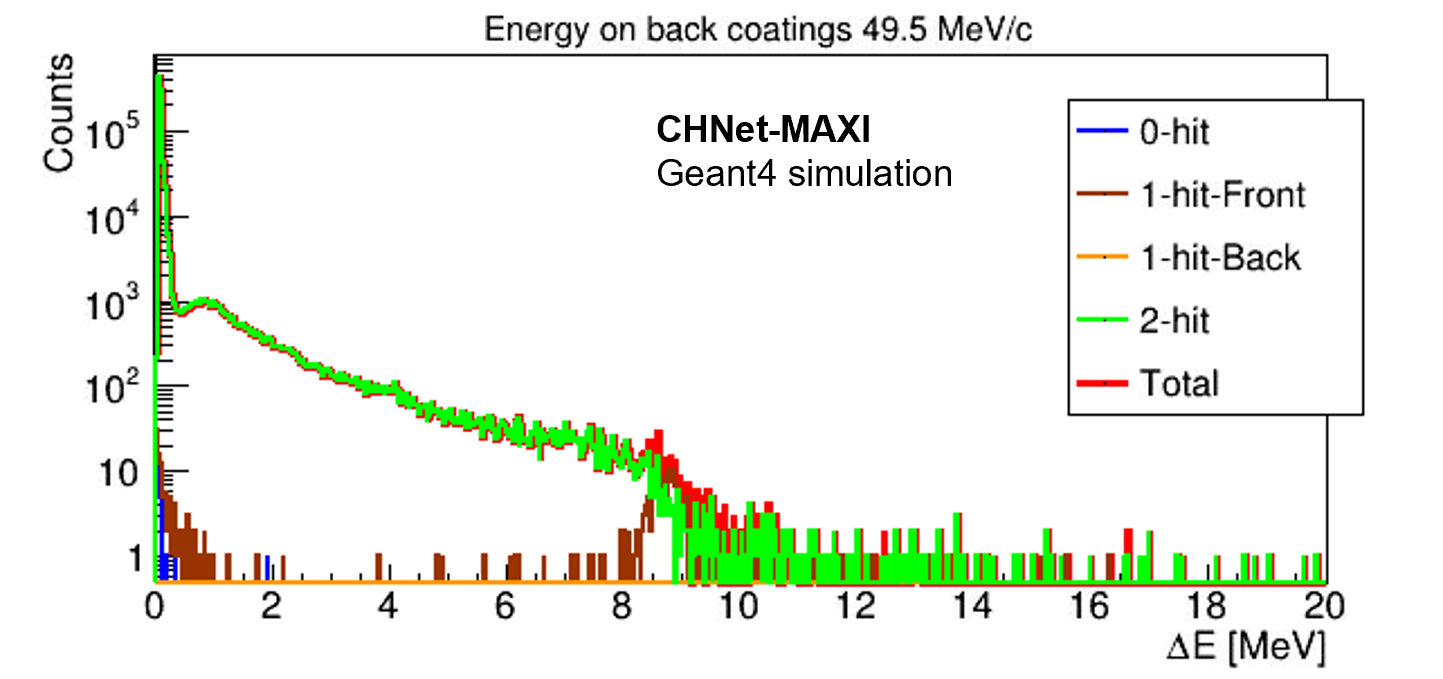}
        \caption{Simulated energy deposit on front (left) and back (right) fibre coatings for 49.5 MeV/c single $\upmu^-$ interacting with 0, 1 or 2 planes of fibres ($10^6$ simulated single muons).}
        \label{fig:sim:Ec}
    \end{figure*}
    \begin{figure*}[p]
        \centering
        \includegraphics[width=0.49\linewidth]{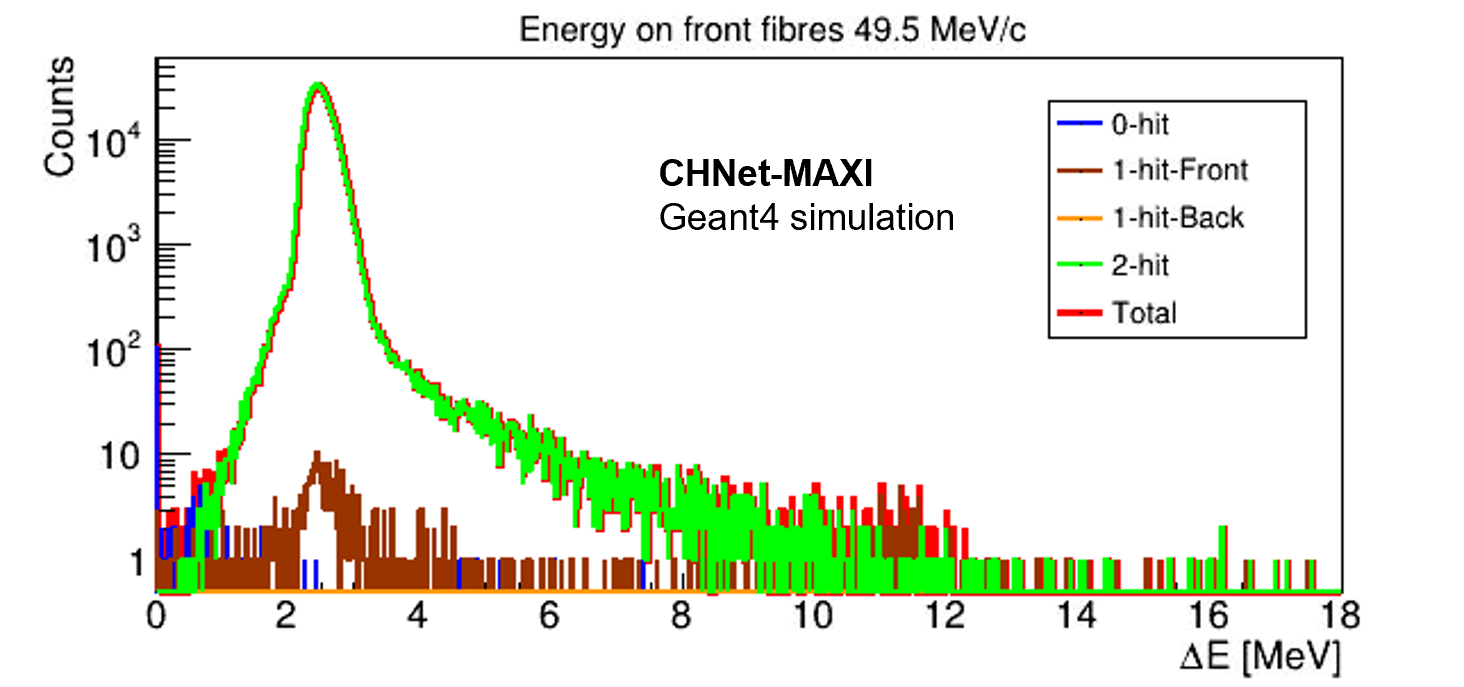}
        \includegraphics[width=0.49\linewidth]{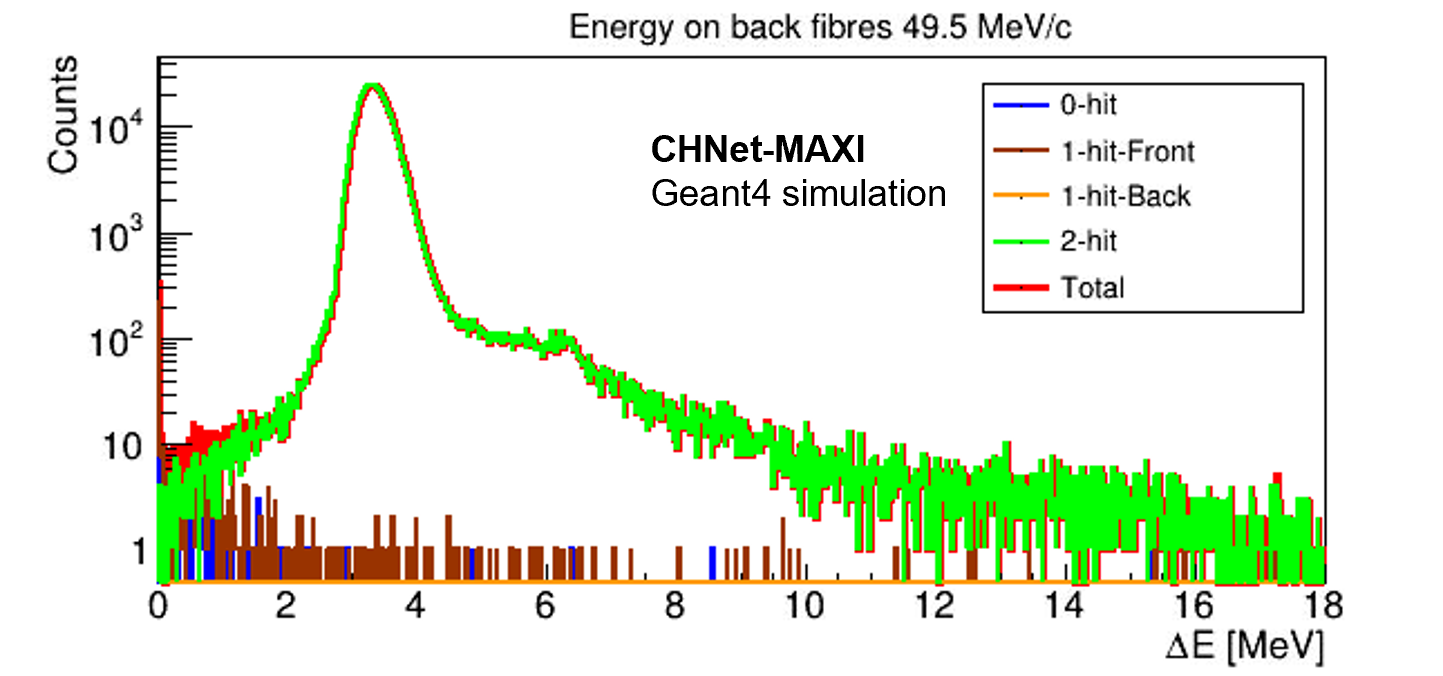}
        \includegraphics[width=0.49\linewidth]{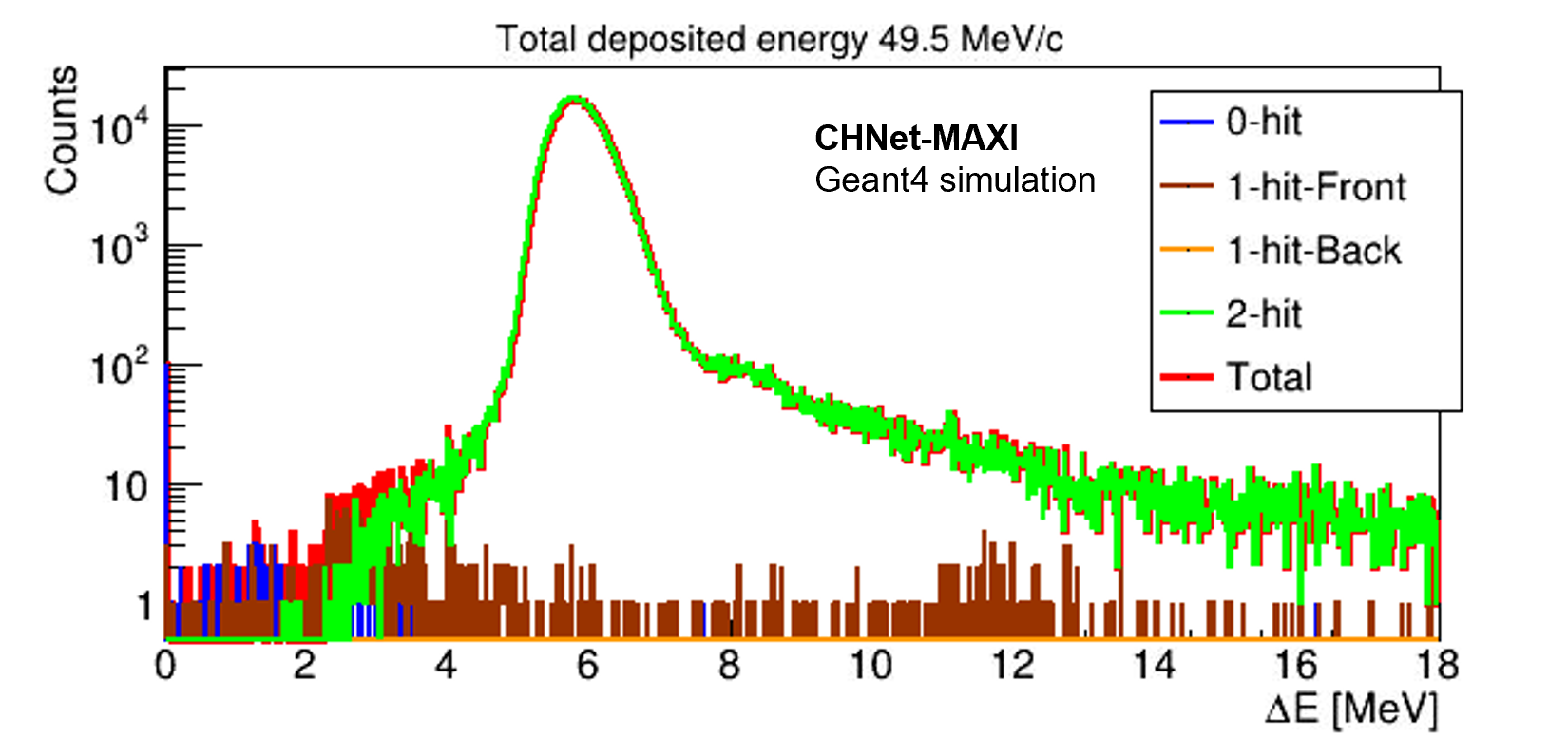}
        \caption{Simulated energy deposit on front (top left), back (top right), and both (bottom) fibre planes for 49.5 MeV/c single $\upmu^-$ interacting with 0, 1 or 2 planes of fibres ($10^6$ simulated single muons).}
        \label{fig:sim:Ef}
    \end{figure*}

    \begin{figure*}[p]
        \centering
        \includegraphics[width=0.4\linewidth]{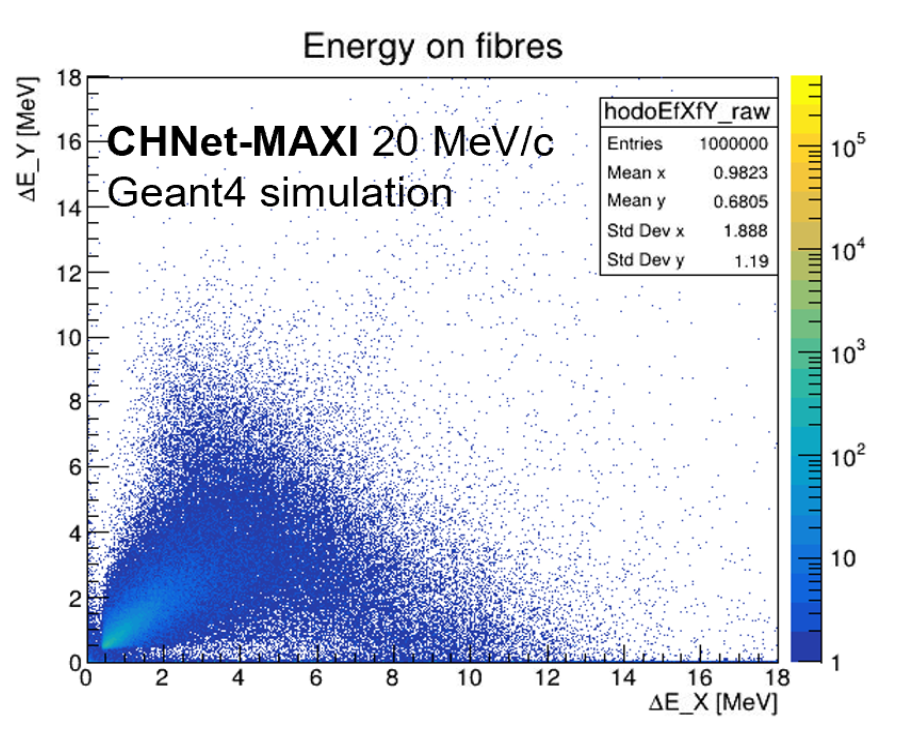}
        \includegraphics[width=0.4\linewidth]{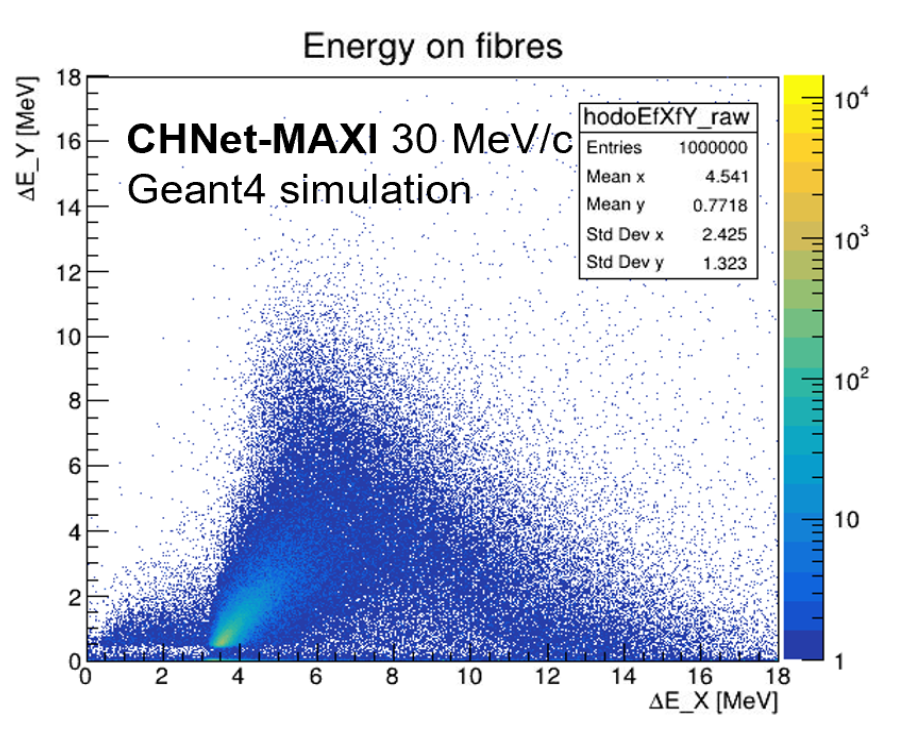}
        \includegraphics[width=0.4\linewidth]{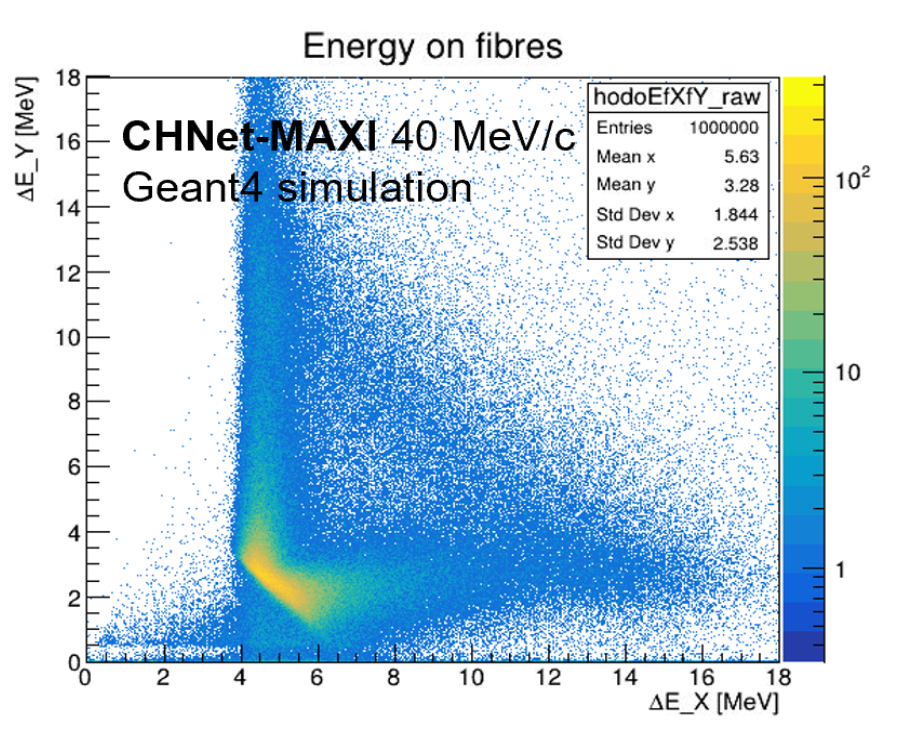}
        \includegraphics[width=0.4\linewidth]{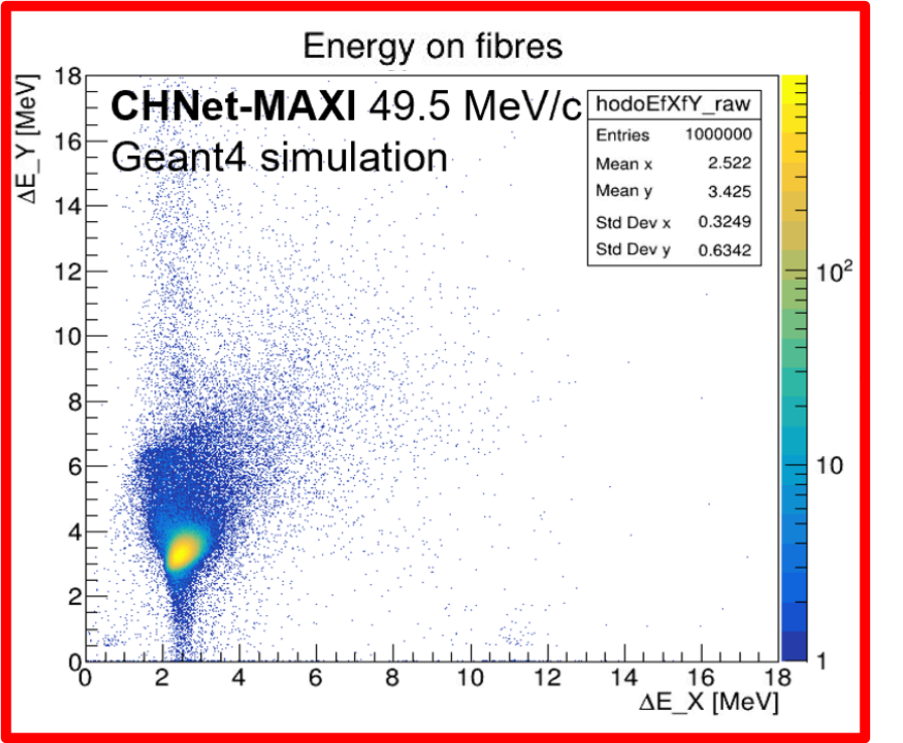}
        \includegraphics[width=0.4\linewidth]{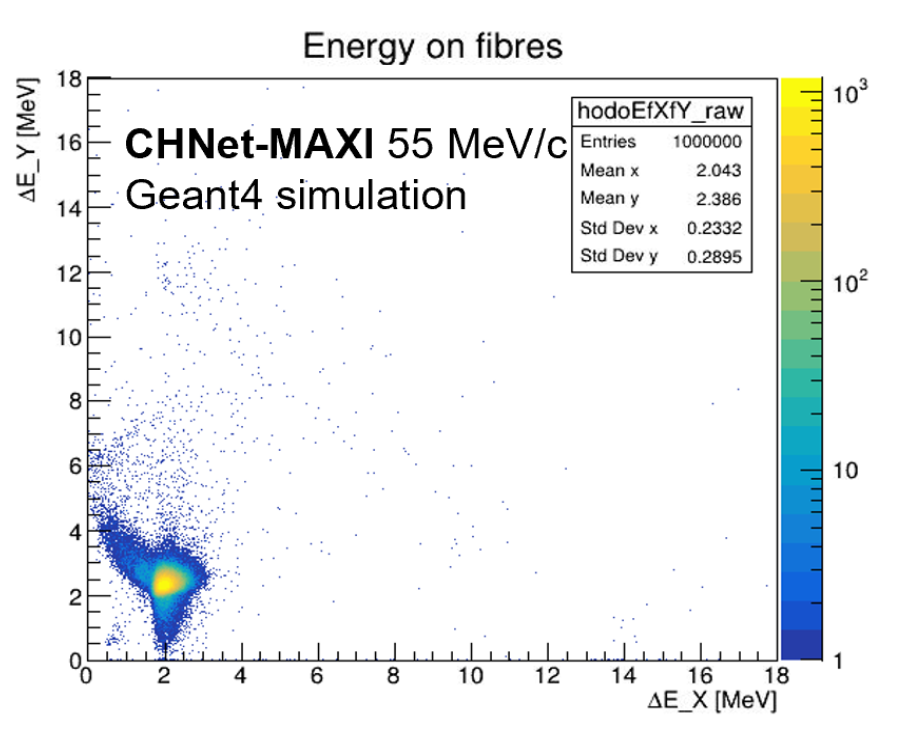}
        \includegraphics[width=0.4\linewidth]{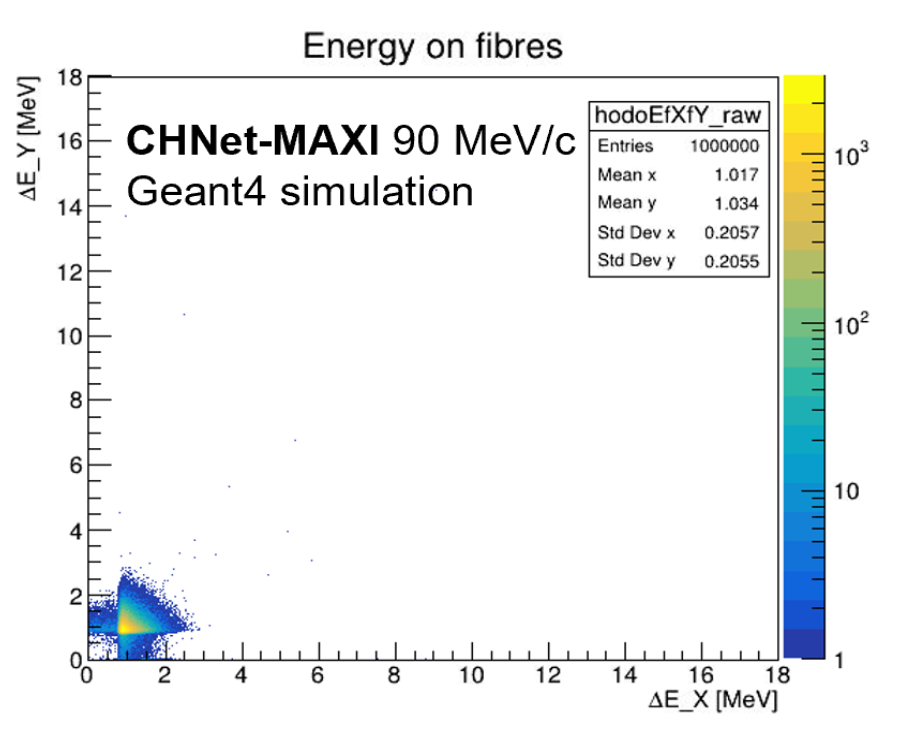}
        \caption{Scatter plot comparing the simulated energy deposit on the two fibre planes ($x$ front, $y$ back) as a function of the negative muon beam momentum, for single negative muons ($10^6$ single muons simulated for each value of momentum). The plot marked in red is the one with the experimental conditions for CHNet-MAXI (49.5 MeV/c $\upmu^-$). At low momentum (20 MeV/c, top left), muons are stopped in the front window and coatings and only secondary particles interact with the fibre; as momentum increases (30 MeV/c, top right, and 40 MeV/c, centre left) muons interact directly with the front ($x$) plane but not with the back one. At 49.5 MeV/c (centre right) muons interact with both planes, releasing much more energy in the back plane as they are close to their range (\emph{i.e.} the maximum of the Bethe-Bloch). As energy increases even more, muons start having an equal response on the two planes (55 MeV/c, bottom left, and 90 MeV/c, bottom right)}
        \label{fig:sim:Ef1Ef2}
    \end{figure*}
    
    The detector has been simulated using a Geant4\cite{agostinelli:geant4} application originally developed for the FAMU experiment muon beam monitor\cite{rossini2024hodo, Rossini:hodosim:2026cimento}. The detector simulation allows us to understand the response of the detector to a beam of muons of the selected energy, and it is also used to compute the geometric coefficients that are part of the detector absolute calibration. 
    
    The geometry consists of 3 mm-pitch polystyrene juxtaposed square fibres, coated with 15 $\upmu$m of TiO$_2$ varnish and arranged in two planes of 32 fibres, as described in Section \ref{sec:hodo}. The two planes are enclosed between two identical 0.5 mm PLA entrance and exit windows. 

    The muon beam simulated for this work is in the 20-90 MeV/c momentum range, which includes the momentum value used for the CHNet-MAXI experiment, \emph{i.e.} 49.5 MeV/c. The energy deposit has been evaluated in all volumes, and considered only for the active volume (the two planes of fibres) for the main purpose of this study. The plots in Figures \ref{fig:sim:Ew}, \ref{fig:sim:Ec}, and \ref{fig:sim:Ef} are control plots and show the energy deposit on the two entrance windows, the fibre coatings, and the fibres, respectively, when the detector is hit with a 49.5 MeV/c negative muon beam. This beam momentum has been chosen in order for muons to stop in the 250 $\upmu$m-thick Pb samples placed just after the beam monitor, meaning that muons in the fibres are almost at the end of range.

    The scatter plots in Figure \ref{fig:sim:Ef1Ef2} show the comparison between the simulated energy deposit on the front ($x$) and back ($y$) fibre planes as a function of the beam momentum for $\upmu^-$. 

    In addition to understanding the detector response, detector simulation plays an important role in its absolute calibration, \emph{i.e.} the conversion of the integrated charge to the number of muons. This activity follows the procedure already carried out for the FAMU beam monitor\cite{rossini2024hodo}. Provided the detector is working in a linear range, far from the SiPM saturation, the beam intensity (particles per second) at a pulsed fixed-energy particle beam evaluated at time $t$ can be written as
    \begin{align}
        \varphi_\mu(t) = \frac{r}{k_g \ Q_\mu}Q_{tot}(t)
    \end{align}
    where $Q_{tot}(t)$ is the instantaneous charge deposit, obtained by summming the integral of the waveforms of all 64 channels, $r$ is the beam pulse rate, $Q_\mu$ is the charge deposited by a muon interacting with both planes of fibres and $k_g$ a geometric correction factor. The latter takes into account the fraction of muons passing through the detector without interacting, or interacting with only one fibre, and it can only be extracted from the simulation. Let's define $W_0$ the fraction of muons not interacting with the fibres (\emph{i.e.} depositing zero energy), and $W_{1f}$ \& $W_{1b}$ the fraction of muons interacting with only one fibre in the front or in the back plane, respectively, therefore depositing a fraction of $Q_\mu$ given by $Q_{1f/b} = Q_\mu \ \eta_{1f/b}$. The geometric factor can therefore be written as:
    \begin{align}\label{eq:kgeo}
        k_g = W_{1f}\ \eta_{1f} + W_{1b}\ \eta_{1b} + W_2 \text{.}
    \end{align}

    \begin{table}[]
        \centering
        \caption{Components of the geometric factor obtained through simulation with statistical uncertainty only ($10^6$ single particles simulated), resulting in a geometric factor $k_g = 0.99946(10)_\text{stat}$, calculated with Equation \ref{eq:kgeo}.}
        \begin{tabular}{c|c|c}
             \toprule
             Fibre hit & $W$ (\%) & $\eta$  \\
             \midrule
             2 & $(99.92 \pm 0.10_\text{stat})$ & 1 \\
             1f & $(0.061 \pm 0.002_\text{stat})$ & $(0.422 \pm 0.008_\text{stat})$\\
             1b & 0 & not def.\\
             0 & $(0.021 \pm 0.002_\text{stat})$ & 0 \\
             \bottomrule
        \end{tabular}
        \label{tab:kgeo}
    \end{table}

    Table \ref{tab:kgeo} reports the values obtained by simulating $10^6$ muons with a geometry similar to the ISIS beam spot at 49.5 MeV/c. The uncertainties provided are only the statistical ones, while further simulation studies are ongoing to extract the systematic uncertainties taking into account the thickness tolerances. The value of geometric factor with statistical uncertainties is $k_g = 0.99946 \pm 0.00010_\text{stat}$, which is about 5 standard deviations from $1$. It is therefore expected for the value of $k_g$ to be consistent with 1 once the correct uncertainty balance is added, \emph{i.e.} no geometric correction would be needed for this detector, thanks to its high packing factor. In this case, the equation to calculate the beam intensity simplifies to
    \begin{align}\label{eq:intensity}
        \varphi_\mu(t) = \frac{r}{Q_\mu} Q_{tot}(t)
    \end{align}
    where $Q_\mu$ will be evaluated through a dedicated low intensity muon beam at 49.5 MeV/c, the momentum chosen for the experiment.

\section{Data acquisition test at CNAO-XPR}

    \begin{figure}
        \centering
        \includegraphics[width=\linewidth]{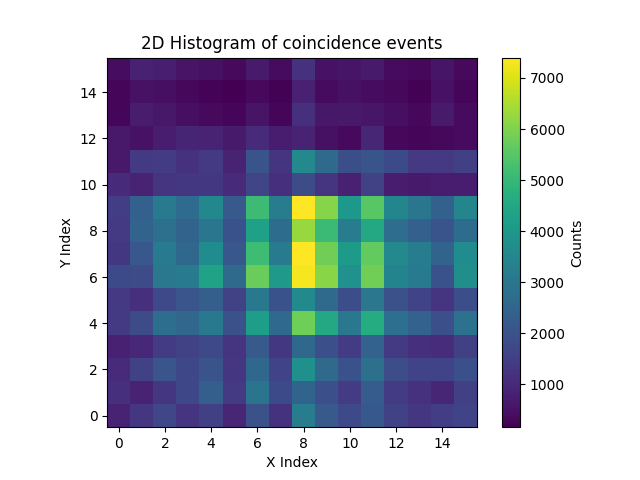}
        \caption{Beam spot with the central 32 channels (16 along $x$ and 16 along $y$) of the detector, obtained with a low intensity 150 MeV proton beam at CNAO. This first operational test proved the detector capability in reconstructing the beam spot, but it also showed the need of an equalisation of the different fibre-SiPM systems.}
        \label{fig:cnao2d}
    \end{figure}
    
    As already done in the past for similar devices\cite{Hodo2}, a first operational test of the detector has been carried out at the National Centre for Oncological Hadrontherapy (CNAO) in Pavia, Italy\cite{cnao1, cnao2}. At this facility, proton and ion beams are directed towards three treatment halls and an experimental room (CNAO-XPR). In particular, the facility is capable of producing proton beams with kinetic energy between 63 MeV and 227 MeV, with a tunable intensity up to a nominal value of $10^{10}$ p/s. 

    The right panel in Figure \ref{fig:hodo} shows the beam monitor set up on the CNAO-XPR beamline with a downstream scintillator used to trigger the DAQ system, similarly to the setup for the FAMU beam monitor test at CNAO\cite{Hodo2}. Only the central 16+16 channels have been connected to the CAEN DT5560SE digitiser, featured with 32 channels sampling at 125 MS/s, 14 bit, with programmable FPGA (OpenFPGA). A custom firmware has been developed using the SciCompiler toolkit by Nuclear Instruments.

    A bias of $+44$ V (about $5.5$ V over the breakdown voltage) was applied to all SiPM boards and to the trigger scintillator SiPMs through a CAEN DT8032M power supply. The trigger scintillator signals were processed through a threshold discriminator set up on one channel of the CAEN DT1081B programmable logic unit, and fed to the digitiser as an external trigger. 
    
    The custom firmware allows both triggered waveform digitisation and waveform integration based on single-particle logic to allow the extraction of $Q_\mu$ to use in equation \ref{eq:intensity}. This logic consists of looking for events in which exactly one fibre per plane had signal corresponding to the beam trigger (operator XOR on fibres of the same plane and operator AND between the two planes).

    For full detector operation, 64 channels are needed, therefore the similar CAEN R5560SE digitiser (128 channels, 125 MS/s, 14 bit) will be operated during final detector deployment. 
    The main reasons for this initial test at CNAO were testing the detector operation and the DAQ capabilities. The detector has therefore been operated at different rates, different beam energies and with the beam spot hitting different positions. 

    Figure \ref{fig:cnao2d} shows a reconstructed 2D beam profile at low intensity, using the XOR-AND condition described in this Section. From this first operation test, it has been clear that an equalisation procedure among the different SiPMs was needed to obtain a uniform spatial response of the detector. This procedure is described in detail in the following Section.

\section{Channel equalisation procedure}
    The equalisation of fibres involves introducing coefficients to the estimation of integral charge of each fibre, such that the same number of photons in each fibre produces the same integrated charge. 

    \begin{figure}[htbp]
        \centering
        \input{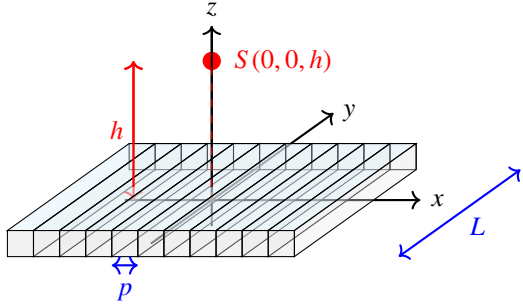}
        \caption{Schematic representation of the fibre plane geometry (squared fibres with pitch $p$ and length $L$) used to calculate the approximate distribution of counts along the fibre number, \emph{i.e.} along the $x$ axis, for a radioactive source $S$ located at a height $h$ above the fibre plane (Equation \ref{eq:final}).}
        \label{fig:scheme}
    \end{figure}

    \begin{figure}
        \centering
        \includegraphics[width=\linewidth]{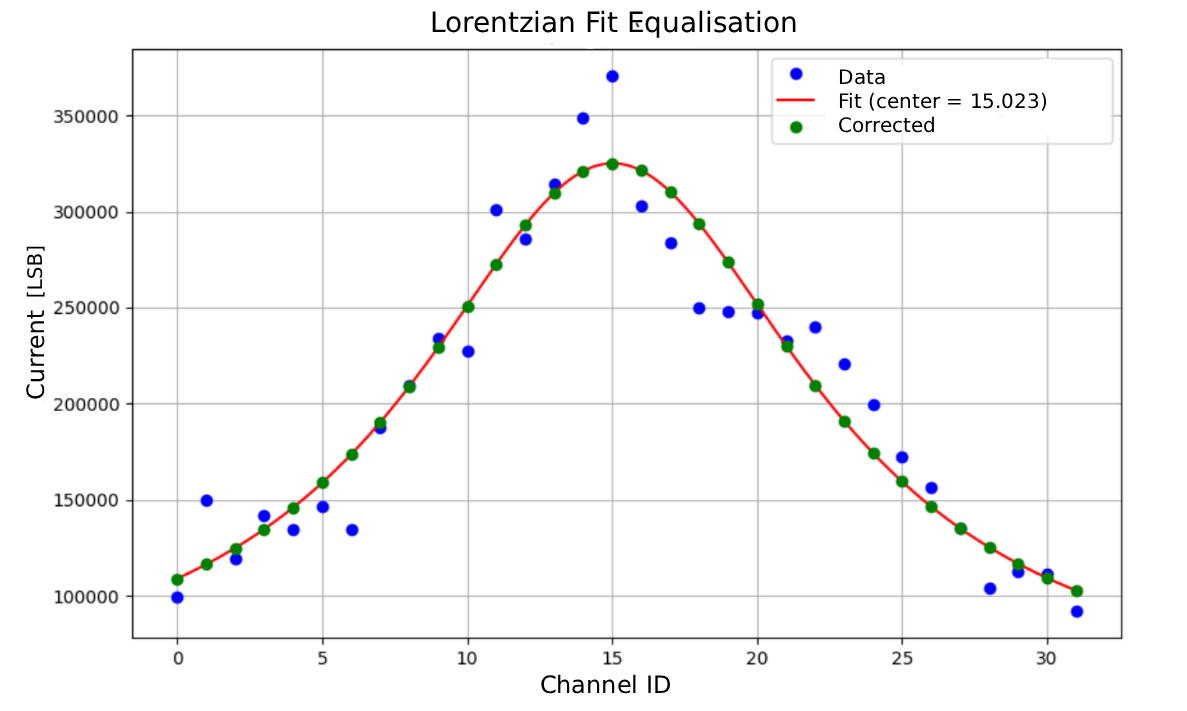}
        \caption{Example of 1D profile on one axis (histogram) acquired with the $^{60}$Co source. The data points are shown in blue, the Lorentz fit in red, and the data sampled from the Lorentz profile in green. The ratio between the measured points and the ones extracted from the fit is used to equalise the channels. The result of equalisation is shown in Figure \ref{fig:eq_spot}.}
        \label{fig:profile}
    \end{figure}

    \begin{figure*}
        \centering
        \includegraphics[width=0.4\linewidth]{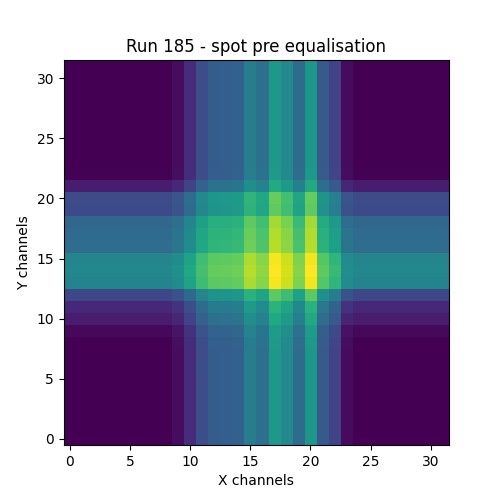}
        \includegraphics[width=0.4\linewidth]{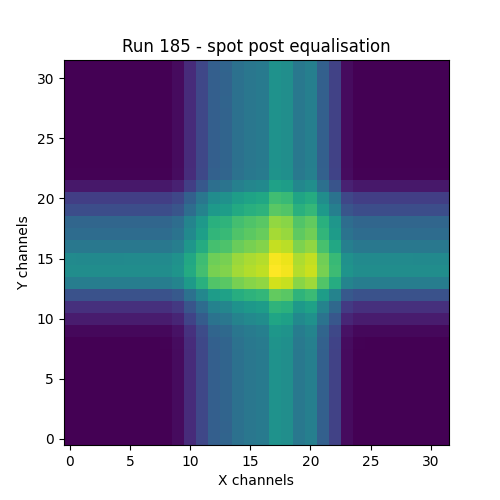}
        \caption{Beam spot projection before (left) and after (right) equalisation procedure, which results in a smoother 2D beam profile. The value in each pixel is obtained from the two 1D profiles as the sum of the corresponding bins on the $x$ and $y$ axes.}
        \label{fig:eq_spot}
    \end{figure*}

    We assume that a point-like isotropic gamma source is positioned at height $h$ over a plane of fibres with length $L$ (along the $y$ axis) and pitch $p$, as in Figure \ref{fig:scheme}, where the source is positioned over the origin of the axes. Assuming the thickness of the fibre is negligible, a fibre element $dA = dx dy$ in position $(x,y)$ is located $R(x,y) = \sqrt{x^2 + y^2 + h^2}$ from the source. Let $\vartheta$ be the angle of sight of the element with respect to the fibre plane, $\cos \vartheta = h/R$, and therefore the number of gammas crossing $dA$ is given by (where $k$ is a constant):
    \begin{equation}
        dN = k \frac{dA \cos \vartheta}{R^2} = kh \frac{dx dy}{R^3(x,y)}
    \end{equation}
    The number of $\gamma$-rays crossing the $i$-th fibre 
    is therefore proportional to:
    \begin{align}
        N_i(x_i) &\propto \int_{x_i-p/2}^{x_i+p/2} dx \int_{y_i-L/2}^{y_i+L/2} dy \ \frac{h}{(x^2 + y^2 + h^2)^{3/2}}
    \end{align}

    Assuming $p \ll h$ (\emph{i.e.} $p \to 0$), the integral over $x$ can be re-written as $\int_{x_i-p/2}^{x_i+p/2} f(x) dx \to p f(x)$. In addition, one can assume $L \gg h$ (\emph{i.e.} $L\to +\infty$), which yields:
    \begin{align} \label{eq:final}
        N_i (x_i) &\propto \int_{-\infty}^{+\infty} dy \ \frac{p h}{(x_i^2 + y^2 + h^2)^{3/2}} \nonumber \\
        &= \left. \frac{p h y}{(x_i^2+h^2)\sqrt{x_i^2 + y^2 + h^2}} \right|_{-\infty}^{+\infty} \nonumber \\
        &= \frac{2p}{h} \ \frac{1}{1+\left(x_i/h\right)^2}
    \end{align}
    which is a Lorentz distribution centred in correspondence to the source and has a Full Width at Half Maximum $\mathtt{FWHM} = 2h$. The known integral $\int \frac{dx}{\left(x^2+a^2\right)^{3/2}} = \frac{x}{a^2\sqrt{x^2+a^2}}+C$ has been used for the calculation.
    

    The derivation shown here is just an approximation of the real case.
    In the experimental situation, $p= 3$ mm, $L=104$ mm and $h = 30$ mm. The source used for the purpose is a $^{60}$Co isotropic source with nominal activity of 0.1 $\upmu$Ci. The $^{60}$Co gamma rays were used because their interactions in the scintillating fibres can produce energy deposits of order 1 MeV, comparable to the energy deposited by a $\sim$50 MeV/c muon crossing a 3-mm-thick plastic scintillating fibre, as discussed in Section \ref{sec:sim}.

    Figure \ref{fig:profile} shows an example of experimental data acquired in these conditions. In particular, it is a histogram of the measured integral on one axis, with the source placed in central position. The fit and the corresponding extrapolated points are also shown. Calling $M_i$ the values measured and $F_i$ the one extrapolated from the fit, the gain correction factor for each $i$-th SiPM has been obtained as $G_i = F_i/M_i$. The gain factors are calculated at different source positions (9 positions have been used), for the twelve fibres around the fit centroid, and the used values are averaged among all quantifications.
    This procedure should be carried out before every experimental campaign of the detector. 

    The resulting gain correction factors calculated with this method range between $0.76$ and $1.46$ across the 64 channels, with a mean of $1.01$ and a standard deviation of $0.12$. This confirms that the correction for single channels is non-negligible. An example of 49.5 MeV/c $\upmu^-$ beam spot measured at RIKEN-RAL Port4 before and after the equalisation procedure is shown in Figure \ref{fig:eq_spot}, showing the procedure was effective in smoothing the beam profile, while keeping the centroid and width stable. This is shown in Table \ref{tab:equalised}, which also shows the fractional roughness, defined to evaluate the discrepancy between a linear approximation among 3 consequent fibres:
    \begin{align}
        R_\text{frac} =& \sqrt{\frac 1 {N-2}\sum_{i=2}^{N-1}\left( \frac{2N_i - N_{i-1}+N_{i+1}}{N_{i-1}+N_{i+1}} \right)^2} \\
        &\text{with } N=32 \text{.}\nonumber
    \end{align}
    As shown in Table \ref{tab:equalised}, the improvement on both $x$ and $y$ axes is measurable, confirming the need and the effectiveness of this calibration procedure.

    \begin{table}[]
        \centering
        \caption{Quality check on the equalisation procedure. Centroid and width appear stable, and they are calculated, respectively, as weighted average and standard deviation of the channels, using the integrated charge as weight. The fractional roughness measures the local linear approximation, it is calculated as  the RMS of the value $r_i =\frac{2N_i - (N_{i-1}+N_{i+1})}{N_{i-1}+N_{i+1}}$ for $i=2,...,31$, and it shows a clear improvement of the signal fluctuations between neighbour fibres.}
        \label{tab:equalised}
        \begin{tabular}{c|c|c|c|c}
            \toprule
             && raw & equalised & $\Delta$\\
             \midrule
             &Centroid $\bar{x}$ (\# fibres) &17.48&17.44 &$-$0.04 \\
             $\mathbf{x}$&Width $\sigma$ (\# fibres)&12.64&12.23&$-$3.2\%\\ 
             &Fractional Roughness $R_\text{frac}$&0.347&0.295&$\mathbf{-15\%}$\\
             \midrule 
             &Centroid $\bar{x}$ (\# fibres) &16.64&16.35 &$-$0.13\\
             $\mathbf{y}$&Width $\sigma$ (\# fibres)&8.17&7.91&$-$3.1\%\\ 
             &Fractional Roughness $R_\text{frac}$&0.334&0.274&$\mathbf{-18\%}$\\

             \bottomrule
        \end{tabular}
    \end{table}

\section{Conclusion and future developments}
    In conclusion, the beam monitor for the CHNet-MAXI experiment described in a previous work\cite{hodomaxi} has been simulated and tested with the CNAO proton beam and with the ISIS RIKEN-RAL negative muon beam. 

    The simulation has been carried out both to understand better the detector response, and to allow a thorough calibration in beam intensity. In particular, the simulation goes in the direction of excluding the need of a geometric factor, simplifying the absolute calibration to Equation \ref{eq:intensity}. The value of the charge deposited by a muon interacting with both planes of fibres is to be extracted during a dedicated low-intensity measurement at RIKEN-RAL using the same momentum as the experiment, 49.5 MeV/c.

    The first test of the detector at CNAO proved the correct operation of this detector and the DAQ system in terms of hardware, firmware, and software, It also showed the need of a gain equalisation procedure in order to get a uniform response from the 64 fibre-SiPM systems.
    The equalisation procedure, carried out with a $^{60}$Co source and through Lorentz fits, allowed the detector to be mounted on the CHNet-MAXI setup at RIKEN-RAL Port4 for the absolute calibration, the last step before the final deployment. It will play a crucial role in the normalisation of the data coming from the X/$\gamma$-ray detectors of the experiment, aiming to show the feasibility of non-destructive isotopic quantification in Pb with Muonic atom X-ray Emission Spectroscopy $\upmu$-XES.

\section*{Acknowledgements}

The CHNet-MAXI experiment has been funded by INFN-CSN5. The INFN Milano-Bicocca and INFN Pavia mechanical workshop, and the INFN Pavia Electronics Service, are gratefully acknowledged for the setting up the beam monitor.
The beamtime at CNAO has been carried out at the CNAO experimental facility (XPR) built in collaboration with INFN.
The beamtime at ISIS has been funded through the 2021-2027 CNR-STFC agreement through approved ISIS experiment nr. RB2400060.

\bibliographystyle{elsarticle-num}
\bibliography{biblio}

\end{document}